\documentclass[journal,article,submit,pdftex,moreauthors]{Definitions/mdpi} 
\firstpage{1} 
\pubvolume{1}
\issuenum{1}
\articlenumber{0}
\pubyear{2022}
\copyrightyear{2022}
\datereceived{} 
\dateaccepted{} 
\datepublished{} 
\hreflink{https://doi.org/}
\preto{\abstractkeywords}{\nolinenumbers}
\pdfoutput=1

\Title{Multifractal cross-correlations of bitcoin and ether trading characteristics in the post-Covid-19 time}
\TitleCitation{Multifractal analysis}

\Author{Marcin Wątorek $^{1,\ddagger}$\orcidA{}, Jarosław Kwapień $^{2,\ddagger}$\orcidB{} and Stanisław Drożdż $^{1,2}$*\orcidC{}}

\AuthorNames{Marcin Wątorek, Jarosław Kwapień and Stanisław Drożdż}

\AuthorCitation{Wątorek, M.; Kwapień, J.; Drożdż, S.}
\address{%
$^{1}$ \quad Faculty of Computer Science and Telecommunications, Cracow University of Technology, ul.~Warszawska 24, 31-155 Krak\'ow, Poland\\
$^{2}$ \quad Complex Systems Theory Department, Institute of Nuclear Physics, Polish Academy of Sciences, Radzikowskiego 152, 31-342 Kraków, Poland}

\corres{Correspondence: marcin.watorek@pk.edu.pl (M.W.)}

\secondnote{These authors contributed equally to this work.}

\abstract{Unlike price fluctuations, the temporal structure of cryptocurrency trading has seldom been a subject of systematic study. In order to fill this gap, we analyse detrended correlations of the price returns, the average number of trades in time unit, and the traded volume based on high-frequency data representing two major cryptocurrencies: bitcoin and ether. We apply the multifractal detrended cross-correlation analysis, which is considered the most reliable method for identifying nonlinear correlations in time series. We find that all the quantities considered in our study show an unambiguous multifractal structure from both the univariate (auto-correlation) and bivariate (cross-correlation) perspectives. We looked at the bitcoin-ether cross-correlations in simultaneously recorded signals as well as in time-lagged signals, in which a time series for one of the cryptocurrencies is shifted with respect to the other. Such a shift suppresses the cross-correlations partially for short time scales, but does not remove them completely. We do not observe any qualitative asymmetry in the results for the two choices of a leading asset. The cross-correlations for the simultaneous and lagged time series become the same in magnitude for the sufficiently long scales.}


\begin{document}

\section{Introduction}

The cryptocurrency market has once again become the media spotlight during a series of spectacular crashes that have driven bitcoin and other principal cryptocurrencies to lose 75-80\% of their market value over roughly half a year till mid-2022 proving that the preceding substantial growths were merely yet another speculation bubble, resembling those that occurred in 2011, 2013, and 2017~\cite{gerlach2018}. This bubble overlapped with the later part of the Covid-19 pandemic~\cite{estrada2020,james2021a,james2021b,kurrey2022,thakur2022}, raising the question of possible relations between the two and opens an interesting topic for future research. It has been shown that the pandemic outburst in early 2020 has some limited effect on the multifractal properties of the price returns~\cite{mnif2020}. The most important observation is that the cryptocurrencies during the outburst lost their potential to be a safe haven, because they started to be strongly cross-correlated with the regular markets~\cite{corbet2020,kwapien2021}. Strength of these correlations has been varying with time, but despite the fact that there happened periods, in which the cryptocurrency market used to evolve rather independently in the post-Covid-19 time~\cite{kwapien2021,allen2022,corbet2022,foroutan2022,mandaci2022}, the overall picture favours strong cross-correlations, especially since Fall 2021. By the post-Covid-19 time we mean a period after the initial pandemic-related panic had simmered down and the markets started to see the pandemic as a part of daily life, i.e., starting from approximately late Spring 2020. Despite the turbulence that the cryptocurrency market has been coping with since the pandemic outburst, there were studies suggesting that it has finally reached maturity and, at least from some perspective, it shows features that are the same as their counterparts in the classic markets~\cite{drozdz2018,drozdz2019}.

Bitcoin (BTC) was introduced in 2009 as the first asset entirely based on the then newly introduced blockchain technology~\cite{nakamoto2008}. Its purpose was to provide a decentralized and non-inflationary alternative for the fiat currencies, which were subject to massive quantitative easing related to the central bank policies aimed at fighting the consequences of the global financial crisis of 2008-2010. At the beginning bitcoin was viewed just as a curiosity even though a transaction that set the bitcoin price expressed in US dollar for the first time took place as early as in May 2010 and the first platform designed for cryptocurrency trading was opened in July 2010~\cite{decker2014}. Soon, the investors found that BTC and other cryptocurrencies, which started to emerge after the idea of decentralized finance gained initial recognition, can be used as speculative assets~\cite{corbet2019}. The first bubble and a subsequent price drop occurred in 2011, when the market was in its infancy stage, while the subsequent bubbles in 2013 and 2017 appeared at later stages of the cryptocurrency market development~\cite{drozdz2018}. This development is characterised from a statistical point of view by a gradual transition from rather idiosyncratic properties of the fluctuation probability distribution functions (pdfs), temporal correlations, and scaling behaviour towards the appearance of the financial stylized facts that are observed in the standard markets~\cite{drozdz2018}.

Nowadays, after 12 years of the existence of this market and an enormous variety of assets traded there, the cryptocurrencies have not yet managed to achieve the initial goal envisaged by their founders. Perhaps the most serious issue that prevents them from being considered as a substitute for the fiat currencies is their extreme volatility, which attracts large amounts of speculative capital, which in turn amplifies price fluctuations. Volatility is one of the standard indicators of asset liquidity: for a liquid asset even a large order does not have any significant impact on its price. However, even the most capitalized cryptocurrencies like BTC or ether (ETH) suffer heavily from large price jumps triggered by large orders. This is related in part to a much smaller trading frequency on various exchanges compared to the stock and foreign currency markets~\cite{makarov2020,fang2022}. Interestingly, this important indicator has seldom been the subject of quantitative studies based on tick-by-tick data. Only recently a progress in this direction has been reported in ref.~\cite{kwapien2022}, where the inter-transaction times, the number of transactions in unit time, and the traded volume have been analysed. Data representing major cryptocurrencies collected from a few trading platforms have shown that the inter-transaction times are long-term autocorrelated with a power-law decay~\cite{kwapien2022} exactly as the data from the regular stock and Forex markets~\cite{ivanov2004,oswiecimka2005,sun2008,kwapien2012,chen2013}. This opens space for a future application and testing of the relevant stochastic models like the Markov switching multifractal duration model~\cite{chen2013,zikes2015} and the continuous-time random walk model~\cite{scalas2000,masoliver2003,gubiec2010,klamut2021} to the cryptocurrency market data. As the inter-transaction times are directly related to the number of transactions in time unit, the latter quantity was also demonstrated to show long-range power-law autocorrelation~\cite{plerou2000}.

The inter-transaction times and the number of transactions in the stock markets were reported to be multifractal~\cite{oswiecimka2005,ruan2011} and the same was observed for the cryptocurrency market data~\cite{kwapien2022}. Multiscaling of the related time series has also been reported with a strong indication that small fluctuations, i.e., the periods of increased trading frequency, show richer multifractality as compared to the large fluctuations associated with the periods of less frequent trading~\cite{kwapien2022}. Slower trading thus happens to be more uncorrelated (efficient) than trading associated with a market frenzy. The long-term autocorrelations of inter-transaction times can also be responsible for their distribution tail behaviour, which according to the ref.~\cite{kwapien2022} cannot be approximated by an exponentially decaying function but, in contrast, in many cases can be approximated by stretched exponential functions or the power-law ones.

In the present work, we study data that represent a few quantities that characterize the trading of two major cryptocurrencies: BTC and ETH. These quantities are: the logarithmic price returns, the volume traded in time unit, and the number of transactions in time unit. We investigate the fractal properties of these data both from the univariate perspective, in which the properties of each signal are analysed separately, and from the bivariate perspective, in which we look at the cross-correlations between the respective time series of BTC and ETH. As a particularly novel element, we also study the lagged cross-correlations between these assets and seek a possible asymmetry between BTC$\rightarrow$ETH and ETH$\rightarrow$BTC directions. Our goal is to answer the following questions:

1. Do the price returns, the volume, and the number of transactions in time unit representing the principal cryptocurrencies show any statistical inter-currency cross-correlations that can be detected with the $q$-dependent detrended cross-correlation coefficient (defined in Section 2)?

2. Are those cross-correlations, if present, fractal? That is, does the covariance of the fluctuations of these quantities reveals multiscaling/multifractality over a range of scales?

3. Do the cross-correlations, if present, survive if the studied signals have been shifted in time with respect to each other?

4. If so, is it possible to observe any asymmetry between the results with respect to a shift direction (BTC$\rightarrow$ETH and ETH$\rightarrow$BTC)?

5. What is a proposed explanation for the outcomes?

Our paper is organised as follows. In Sect.~\ref{sect::methods} we present the data on which our study is based together with the applied multifractal formalism. In Sect.~\ref{sect::results} we report details of the results and in Sect.~\ref{sect::discussion} we sum up the results and discuss their implications.

\section{Materials and Methods}
\label{sect::methods}

In this study, we analyse a high-frequency data collected from Binance~\cite{binance}, which is the largest cryptocurrency trading platform in terms of daily volume~\cite{coinmarketcap}. We analyse time series representing the number of transactions in time unit $N(t_i)$, logarithmic price returns $r(t_i)$, and volume traded in time unit $V(t_i)$ for two major cryptocurrencies: BTC and ETH, which were sampled every $\Delta t=10$s with $\Delta t = t_{i+1}-t_i$ and $i=1,...,T$. Since the cryptocurrency trading is continuous 24/7, our time series that start on Apr 1, 2020 and end on May 31, 2022 (i.e., they cover the period that we call the post-Covid-19 one) consist of 791 trading days and their length equals $T=6,834,240$ data points.

Among a few available approaches to fractal analysis of time series, the multifractal detrended fluctuation analysis (MFDFA) has proven to be among the most reliable (see, e.g.,~\cite{oswiecimka2006}). The reliability of MFDFA was assessed by comparing its outcomes for a few model data sets with the respective theoretical values based on analytically derived formulas. The MFDFA performance was in majority of cases much better than that of competitive methods. MFDFA was designed to deal with nonstationary data by independently removing trends on different time scales and to examine the statistical properties of the residual fluctuations~\cite{kantelhardt2002}. Its generalized version, the multifractal cross-correlation analysis (MFCCA~\cite{oswiecimka2014}), is capable of detecting multiscale cross-correlations between two parallel nonstationary signals~\cite{podobnik2008,zhou2008,horvatic2011}. Here, we briefly sketch the MFCCA procedure.

Let us consider two nonstationary time series ${\rm X}=\{X_i\}_{i=1}^T$ and $Y=\{Y_i\}_{i=1}^T$ of length $T$ sampled uniformly with an interval $\Delta t$. We start by dividing each time series into $M_s=2 \lfloor T/s \rfloor$ disjoint segments of length $s$ going from both its start ($i=1$) and its end ($i=T$), where $\lfloor \cdot \rfloor$ denotes the floor value. Next, we integrate the time series within each segment $\nu$ and remove a polynomial trend $P_{\nu}^m(j)$ of degree $m$ from the resulting integral signal:
\begin{equation}
x_j(s,\nu) = \sum_{k=1}^j X_{j(\nu-1)+k} - P_{\nu}^m(j), \quad j=1,...,s, \quad \nu=1,...,M_s.
\label{eq::detrended.profile}
\end{equation}
Typically, a polynomial of degree $m=2$ is used, because the results obtained for a few larger values of $m$ were stable. In each segment, we calculate a detrended covariance:
\begin{equation}
f_{\rm XY}^2(s,\nu) = {1 \over s} \sum_{j=1}^s \left[ x_j(s,\nu) - \langle x_j(s,\nu) \rangle_j \right] \left[ y_j(s,\nu) - \langle y_j(s,\nu) \rangle_j \right],
\label{eq::detrended.covariance}
\end{equation}
where $\langle \cdot \rangle_j$ denotes the averaging over $j$. We then use the covariances of all segments to calculate the signed moments of order $q$, which are called the (bivariate) fluctuation functions of $s$:
\begin{equation}
F_q^{\rm XY}(s) = \big\{ {1 \over M_s} \sum_{\nu=1}^{M_s} {\rm sign} [f_{\rm XY}^2(s,\nu)] | f_{\rm XY}^2(s,\nu)|^{q/2} \big\}^{1/q}.
\label{eq::fluctuation.function.xy}
\end{equation}
Since covariances can be negative, their absolute value prevents $F_q^{\rm XY}$ from being complex, while the sign function ensures the consistency of the results~\cite{oswiecimka2014}. A possibly negative sign of the whole expression in the curly brackets also has to be preserved before the $q$th-degree root is calculated. A character of the functional dependence of $F_q^{\rm XY}$ on $s$ allows for distinguishing the fractal time series from those that do not show this property. The most interesting are those signals for which the fluctuation functions exhibit power-law scaling for some range of $q$ and $s$:
\begin{equation}
F_q^{\rm XY}(s) \sim s^{\lambda(q)},
\end{equation}
where $\lambda(q)$ plays a role of the bivariate generalized Hurst exponent. If the cross-correlations are monofractal, then $\lambda(q)={\rm const}$ for all $q$. In contrast, a multifractal case is associated with a monotonically decreasing function $\lambda(q)$.

A special case of $F_q^{\rm XY}$ is X=Y, when the detrended cross-correlations become the detrended autocorrelations. In this case, we can omit both the sign term and the modulus in Eq.~(\ref{eq::fluctuation.function.xy}) as the detrended variance $f_{\rm X}^2$ is always positive. The MFCCA then reduces to the standard MFDFA with univariate fluctuation functions $F_q^{\rm XX}(s)$ and $F_q^{\rm YY}(s)$. The detrended cross-correlation function plays a role of a mean covariance, while $F_q^{\rm XX}$ and $F_q^{\rm YY}$ play a role of mean variances. If the univariate fluctuation functions are power-law dependent on $s$ such that $F_q^{\rm XX}\sim s^{h_{\rm X}(q)}$ and $F_q^{\rm YY}\sim s^{h_{\rm Y}(q)}$, the exponents $h_{\rm X}(q)$ and $h_{\rm Y}(q)$ are the generalized Hurst exponents, which for $q=2$ reduce to the standard definition of the Hurst exponent $H$.

Having calculated all fluctuation functions, we can then introduce the $q$-dependent detrended cross-correlation coefficient $\rho_q(s)$~\cite{kwapien2015} defined as
\begin{equation}
\rho_q(s) = {F_q^{\rm XY}(s) \over \sqrt{ F_q^{\rm XX}(s) F_q^{\rm YY}(s) }}.
\label{eq::rho.q}
\end{equation}
Formula~\ref{eq::rho.q} was proposed in such a form in order to resemble the formula for the Pearson cross-correlation coefficient if $q=2$. The coefficient $\rho_q(s)$ can be considered then as a counterpart of the Pearson coefficient for non-stationary signals. Both coefficients assume values in the range [-1,1] with $\rho_q(s)=1$ for perfectly correlated time series, $\rho_q(s)=0$ for independent time series, and $\rho_q(s)=-1$ for perfectly anticorrelated time series. It should be noted that, in order to calculate $\rho_q(s)$ the time series do not have to be fractal~\cite{kwapien2015}.

If compared with the Pearson coefficient, the coefficient $\rho_q$ offers a few main advantages. The first one is its flexibility of the trend removal. There is no a priori the best polynomial to use in Eq.~(\ref{eq::detrended.profile}). The order $m$ of this polynomial has to be optimised by considering the stability of the final results if we vary $m$. The lowest order that gives stable results is preferred. Typically, the optimal value of $m$ is larger than 1, so the removed trends can be nonlinear, which is important especially for the large scales $s$. This contrasts with more a standard approach to detrending of the financial data, where the linear trends are considered (e.g. by transforming the original time series to its increments). In general, it is also possible to use $m$ that is selected individually for each segment, but we shall not consider such a detrending variant here. Another advantage of $\rho_q$ over the Pearson coefficient is that it is not sensitive to the linear correlations only -- we can look at its values for different $q$s and catch the nonlinear correlations as well. Moreover, with $\rho_q$ we can have some insight into the amplitude of the fluctuations that carry the correlations (by tuning $q$ we amplify the segments of specific variance/covariance). The fourth advantage is that $\rho_q$ is inherently multiscale-oriented. In order to achieve a comparable feature by using the Pearson coefficient one needs to consider the same observable recorded with different sampling frequencies, while for $\rho_q$ it is a built-in property. The fifth advantage is that the DFA procedure and its multifractal generalisations are able to detect the fractal and multifractal scaling in the data fluctuations, something that is beyond the scope of the Pearson-coefficient-based correlation measures. Detection of the multifractality of data allows one to gain some insight into the nature of the processes that govern the observed time series evolution. It can be of importance if one's goal is to model the empirical data in order to make prediction, for example. All these properties that we have mentioned here make the use of $\rho_q$ highly recommended even if the computation in this case is more resource-demanding than in the case of the standard correlation coefficient.

\section{Results}
\label{sect::results}

\begin{figure}[H]
\includegraphics[width=1.0\textwidth]{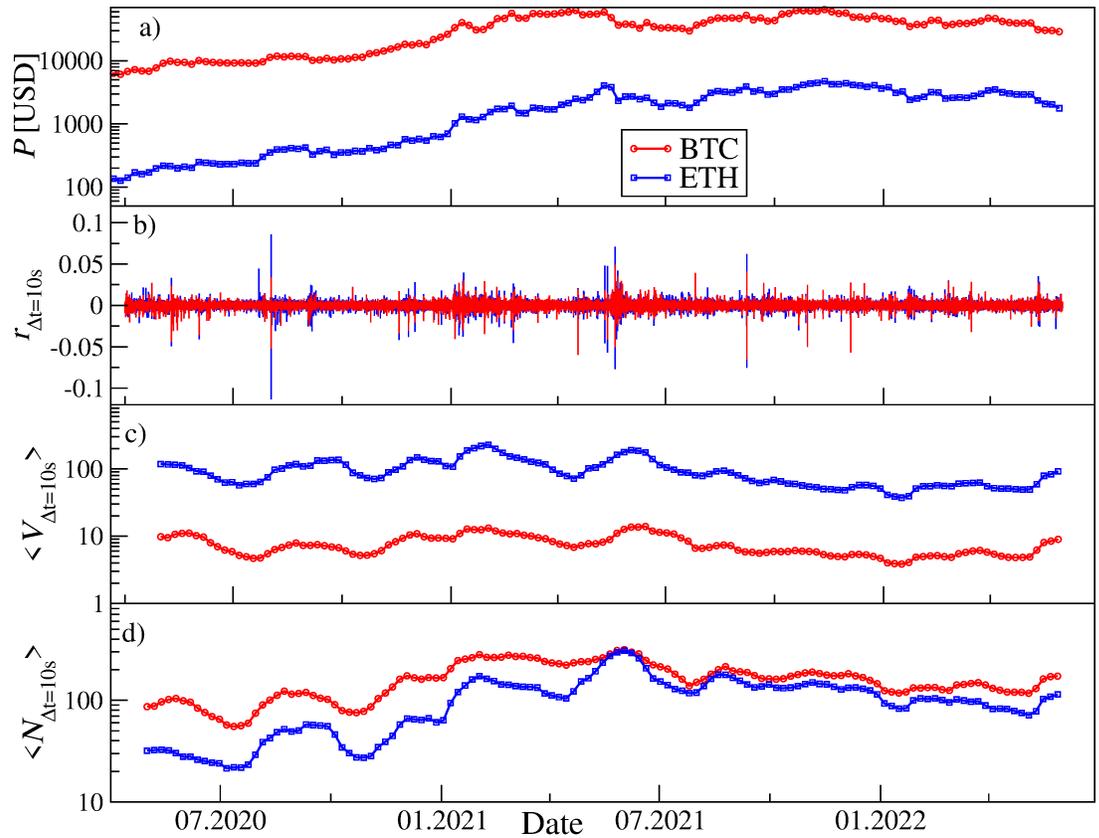}
\caption{Evolution of the quantities of interest over the time period considered in this study for two principal cryptocurrencies: BTC (red circles) and ETH (blue squares). (Top) Price $P(t)$ of the cryptocurrencies expressed in US dollars; (upper middle) logarithmic returns $r_{\Delta t}(t)$ for $\Delta t=10$s; (lower middle) mean volume traded $ \langle V_{\Delta t}(t) \rangle$ in 10s intervals; (bottom) mean number of transactions $\langle N_{\Delta t}(t) \rangle$ in $\Delta t=10$s. The averaging was carried out over a rolling window of 1 month with a step of 6 days.}
\label{fig::temporal.evolution}
\end{figure}

We begin the presentation by taking a look at the time evolution of the quantities of interest over the interval considered in this work. Fig.~\ref{fig::temporal.evolution} (top) shows the price course of BTC and ETH expressed in tether (USDT) -- a stablecoin pegged (1:1 on average) to the US dollar~\cite{tether}. This stablecoin is used by trading platforms as a proxy for the fiat currencies, because it facilitates trading and allows the market participants for avoiding taxation while temporarily closing the cryptocurrency positions. The lowest prices of BTC and ETH during the observed 26 months were 6,202 USDT and 130 USDT, respectively, recorded on Apr 1, 2020 and the highest prices were 68,789 USDT recorded for BTC on Nov 10, 2021 and 4,892 USDT for ETH recorded 6 days later. The time span considered comprised almost the entire rally and subsequent fall of this market.

The logarithmic returns presented in Fig.~\ref{fig::temporal.evolution} (upper middle) are defined as
\begin{equation}
r_{\Delta t}(t_i) = \ln P(t_{i+1}) - \ln P(t_i),
\end{equation}
where $P(t_i)$ is an asset price at time $t_i$. Inferring their heavy-tailed probability distribution function and volatility clustering is straightforward. Periods of market turbulence are associated with high amplitude of the returns. In Fig.~\ref{fig::temporal.evolution} a rolling window of 1 week is applied to calculate the mean volume traded $\langle V_{\Delta t}(t) \rangle$ (lower middle), and the mean number of transactions executed in 10s-long intervals (bottom). The market sensed the most increased volatility level in September 2020, January-February 2021, and May-June 2021. These volatile periods can be associated with a few bear phases of the cryptocurrency market. The periods of increased volatility overlap with the periods of increased volume and increased number of transactions, as can be seen in Fig.~\ref{fig::temporal.evolution}.

\begin{figure}[H]
\includegraphics[width=1.0\textwidth]{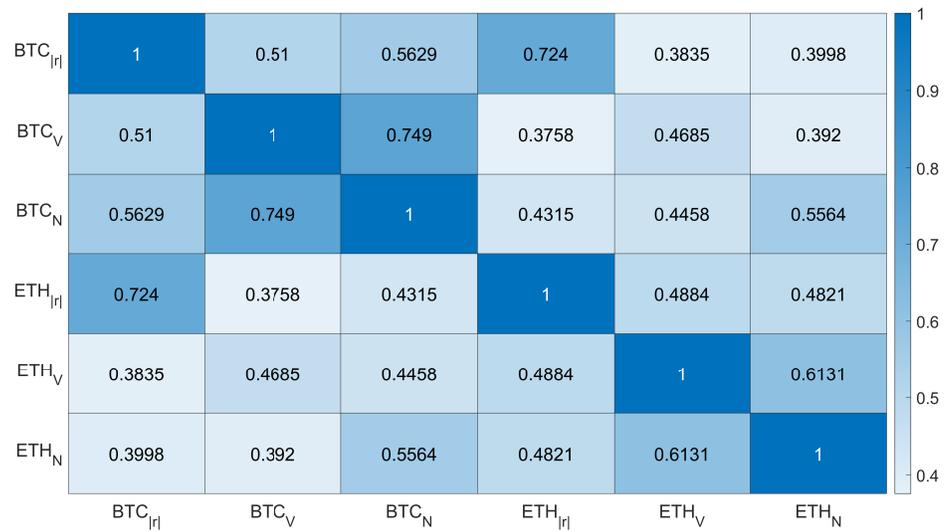}
\caption{Pearson's cross-correlation coefficients calculated for all possible pairs of time series considered in this study. All values are statistically significant.}
\label{fig::Pearson.coefficient}
\end{figure}

These associations can be quantified in terms of the Pearson cross-correlation coefficient
\begin{equation}
C_{\rm XY}= {(1/T) \sum_{i=1}^T \left(X_i - \langle X \rangle \right) \left(Y_i - \langle Y \rangle \right) \over \sigma_{\rm X} \sigma_{\rm Y}},
\label{eq::Pearson}
\end{equation}
where $\langle \cdot \rangle$ is mean and $\sigma$ is standard deviation of the time series $\{X_i\}$ and $\{Y_i\}$. Values of $C_{\rm XY}$ for all pairs of the non-detrended time series are collected in Fig.~\ref{fig::Pearson.coefficient} showing that the strongest cross-correlations are observed for the absolute values of logarithmic returns of BTC and ETH ($C_{\rm XY}=0.72$) and for the mean number of transactions and the mean volume traded of BTC ($C_{\rm XY}=0.75$). All the related time series representing BTC are substantially cross-correlated among themselves, and the same can be noted regarding the time series representing ETH. The least cross-correlated are the pairs in which time series represent both cryptocurrencies, but even in this case, the obtained values are statistically significant.

\begin{figure}[H]
\includegraphics[width=1.0\textwidth]{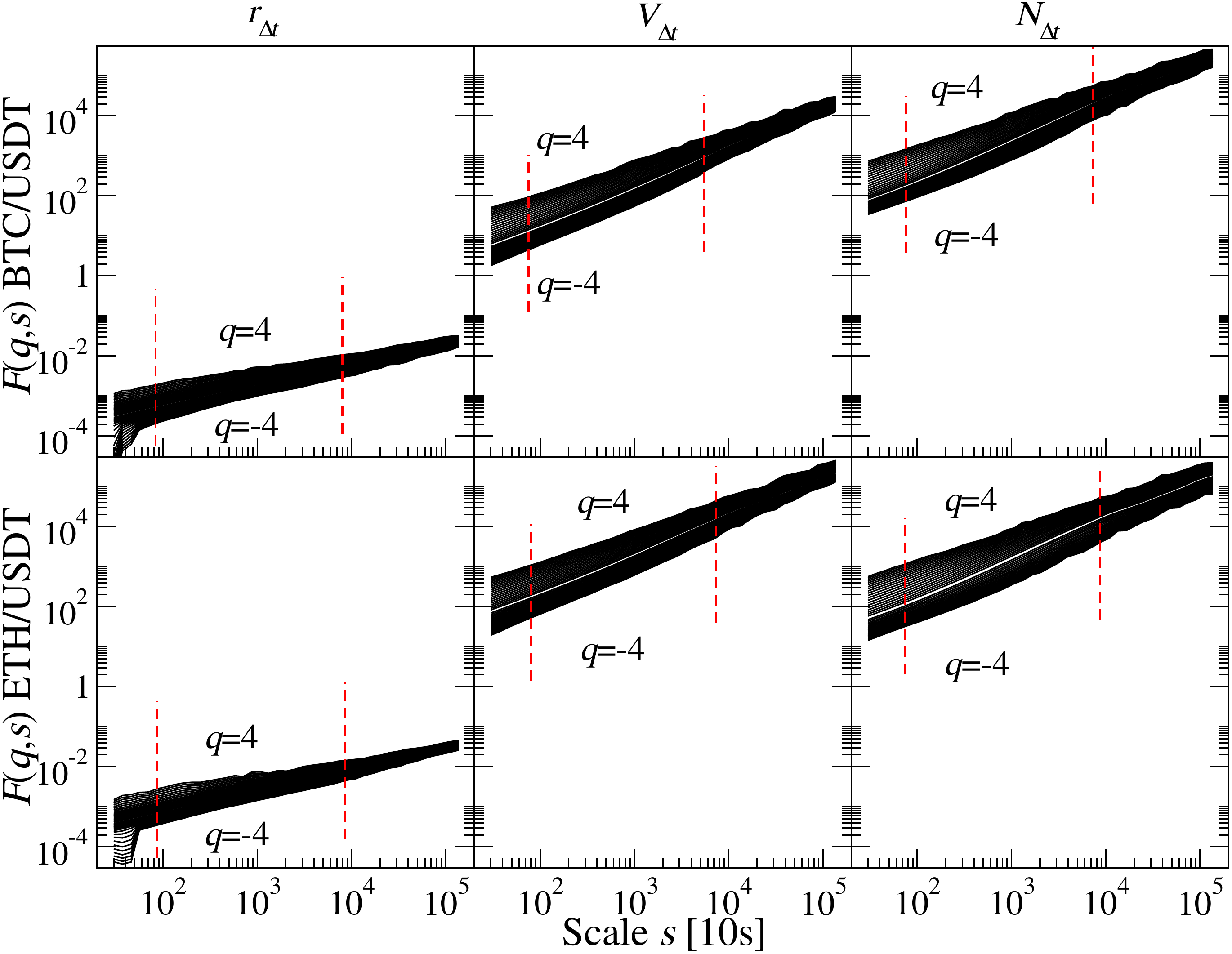}
\caption{Univariate fluctuation functions $F_q^{\rm XX}(s)$ calculated for time series of price returns $r_{\Delta t}(t)$ (left column), volume traded $V_{\Delta t}(t)$ (middle column), and the number of transactions $N_{\Delta t}(t)$ (right column) for two cryptocurrencies expressed in USDT: BTC (top) and ETH (bottom). In each panel, vertical dashed lines denote a range of time scales $s$ for which a power-law model can be fitted to the fluctuation functions. A range of values of $q$ is also shown.}
\label{fig::fluctuation.function.auto}
\end{figure}

Now we pass on to an analysis of the detrended time series. First, let us examine the fractal autocorrelations in terms of univariate fluctuation functions $F_q^{\rm XX}(s)$. The results for all time series of interest are plotted in Fig.~\ref{fig::fluctuation.function.auto}. All plots exhibit a clear scaling for 2-3 decades for a range of $q$, which indicates that the time series are fractal. Moreover, a 'broom'-like shape of the function plots for different values of $q$ may be interpreted as a signature of the multifractal nature of time series. This result goes in parallel to that previously reported for the mean number of transactions executed on different trading platforms~\cite{kwapien2022}. For $q=2$ we obtain the Hurst exponent $H$ that serves as a measure of long-term autocorrelations. One can notice that the time series for $V_{\Delta t}$ and $N_{\Delta t}$ are more persistent (a steeper ascent of $F_{q=2}^{\rm XY}(s)$) than the time series for $r_{\Delta t}$ (a milder ascent).

\begin{figure}[H]
\includegraphics[width=1.0\textwidth]{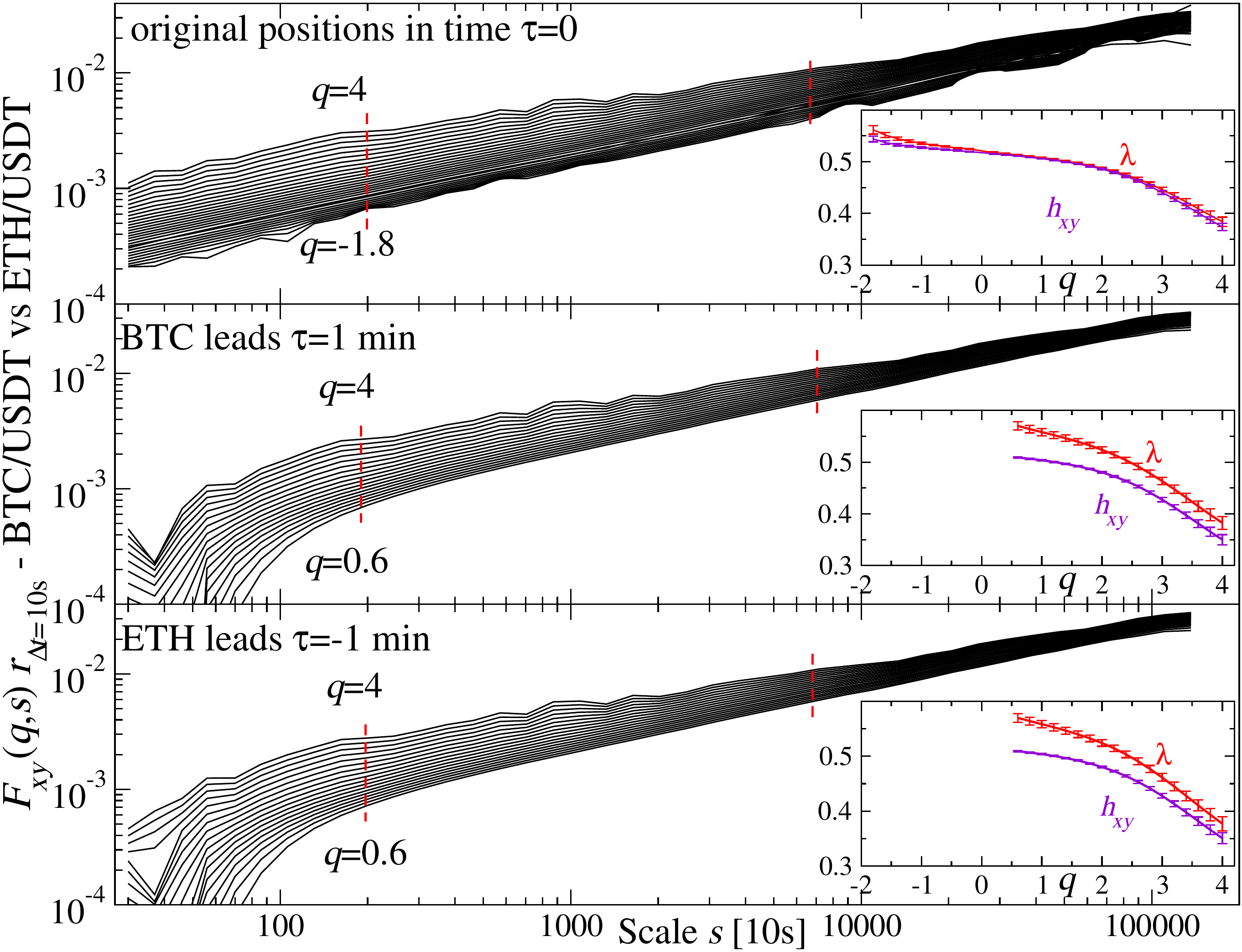}
\caption{(Main panels) Fluctuation functions $F_q^{\rm XY}(s)$ calculated for the time series of logarithmic price returns $r_{\Delta t}(t)$ for X=BTC and Y=ETH. In each panel, vertical dashed lines denote a range of time scales $s$ for which a power-law model can be fitted to $F_{\rm XY}^q(s)$. Extreme values of $q$ are also shown. Three cases are considered: both time series are simultaneous (top), time series representing BTC is advanced by $\tau=1$min (middle), and time series representing ETH is advanced by $\tau=1$min. (Insets) The bivariate scaling exponent $\lambda(q)$ vs. the mean univariate scaling exponent $h_{\rm XY}(q)$ calculated for the same time series. Error bars denote the standard errors.}
\label{fig::fluctuation.function.cross.returns}
\end{figure}

\begin{figure}[H]
\includegraphics[width=1.0\textwidth]{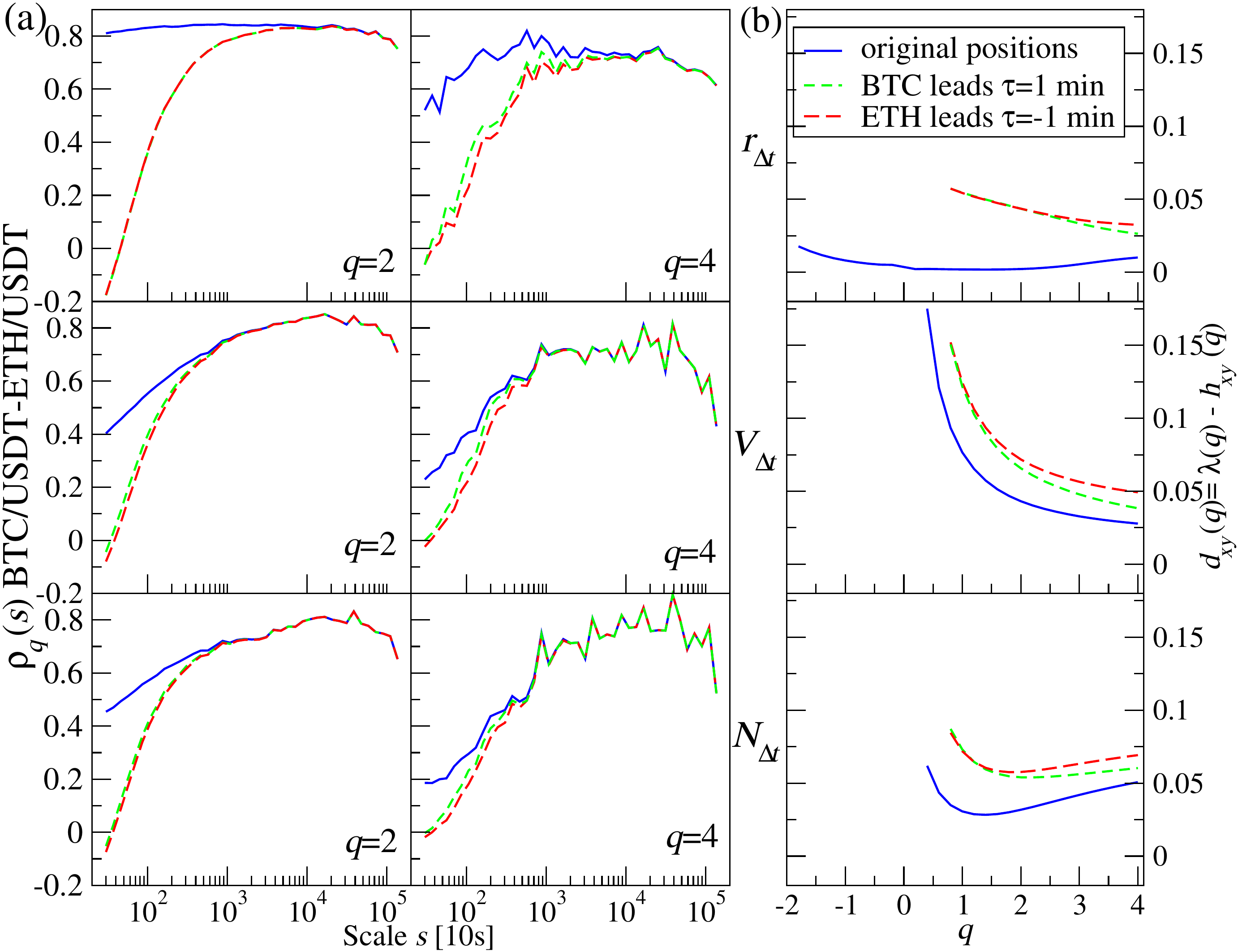}
\caption{(a) The $q$-dependent detrended cross-correlation coefficient $\rho_q(s)$ for $q=2$ (left) and $q=4$ (right) calculated for time series of price returns $r_{\Delta t}(t)$ (top), volume traded $V_{\Delta t}(t)$ (middle), and the number of transactions $N_{\Delta t}(t)$ in $\Delta t$ (bottom) for X=BTC and Y=ETH. Three cases are considered: both time series are simultaneous (solid blue), time series representing BTC is advanced by $\tau=1$min (short-dashed green), and time series representing ETH is advanced by $\tau=1$min (long-dashed red). (b) Difference $d_{\rm XY}(q)$ between the bivariate scaling exponent $\lambda(q)$ and the mean univariate exponent $h_{\rm XY}(q)$.}
\label{fig::rho.q}
\end{figure}

Encouraged by the values included in Fig.~\ref{fig::Pearson.coefficient} and our previous results on the price returns in the pre-Covid-19 era~\cite{watorek2021}, which proved to be fractally cross-correlated, we now study the detrended cross-correlations between the time series representing BTC/USDT and ETH/USDT cross-rates. Fig.~\ref{fig::fluctuation.function.cross.returns} (top) shows the bivariate fluctuation functions obtained for three particular time series arrangements. The original time series, parallel in time, develop $F_q^{\rm XY}(s)$ that scales over $\sim 2.5$ decades for $-1.8 \le q \le 4$, which is quite extraordinary, because cross-correlations typically only scale for positive q. The insets in Fig.~\ref{fig::fluctuation.function.cross.returns} show the $q$-dependence of the bivariate scaling exponent $\lambda(q)$ and the mean univariate one: $h_{\rm XY}(q) = (1/2) [h_{\rm X}(q) + h_{\rm Y}(q)]$. Both are decreasing functions of $q$, which is a signature of multiscaling, and for $-2 \le q \le 4$ they are equal up to their standard errors. This equality suggests that there is little difference in the scaling properties of both time series. If we look at this gap as a separate quantity $d_{\rm XY}(q)=\lambda(q)-h_{\rm XY}(q)$, we see that for $q \ge 2$ it starts to increase again slightly -- Fig.~\ref{fig::rho.q}(b) (blue line in top panel). A relative behaviour of $\lambda(q)$ and $h_{\rm XY}(q)$ is associated with the coefficient $\rho_q(s)$ in such a way that if $\lambda(q) > h_{\rm XY}(q)$ for a given $q$, then $\rho_q(s)$ increases with $s$ the more the larger is this difference. Consequently, the convergent exponents indicate that this increase in the detrended cross-correlations loses momentum with increasing $q$, while the divergent ones indicate the opposite~\cite{kwapien2015}. Indeed, the coefficient $\rho_q(s)$ shows steeper growth for $q=4$ than for $q=2$ in the top panels of Fig.~\ref{fig::rho.q}(a) (blue lines). For $q=2$ it is even roughly constant with only minor decrease for the shortest scales and the longest ones.

The middle and bottom panels of Fig.~\ref{fig::fluctuation.function.cross.returns} present $F_q^{\rm XY}(s)$ for the time series that are shifted relative to each other by $\tau=1$ min: either BTC leads by $\tau$ (middle) or ETH leads by $\tau$ (bottom). In both cases, we see similar power-law dependence over $\sim 2$ decades, but for larger $s$ than in the $\tau=0$ case. We consider this shift toward longer $s$ as expected, because asset prices need time to build up the cross-correlations if they are weakened by the relative shifts. The bivariate and mean univariate scaling exponents shown in the insets reveal a gap between them that is the largest for $q<1$, while if $q$ increases, both quantities gradually converge. Compare this with Fig.~\ref{fig::rho.q}(b) (top panel, green and red lines) for an even better visibility. The plots for $\tau=0$ and $\tau=\pm1$ differ clearly from each other: for the shifted time series $\rho_q(s)$ is more variable than for the simultaneous ones. However, what is similar is that here $\rho_q(s)$ also shows steeper growth for $q=2$ than for $q=4$ -- see Fig.~\ref{fig::rho.q}(a) (top panels, green and red lines). For $q=2$ there is no visible difference between the results if either BTC or ETH leads (top left). This changes after we move to higher $q$s: for $q=4$ there is a noticeable difference between the corresponding lines with higher values of $\rho_q(s)$ if BTC leads. This difference, however, is much smaller that that between the results for $\tau=0$ and $\tau=\pm 1$. This result is somehow expected, because contemporary markets operate at scales that are much shorter than 1 minute. However, due to the long-range temporal autocorrelation in volatility lasting up to a few trading days~\cite{watorek2021} even if we shift one time series with respect to the other by 1 min, the fractal structure of the detrended cross-correlations can still be observed for these time series.

\begin{figure}[H]
\includegraphics[width=1.0\textwidth]{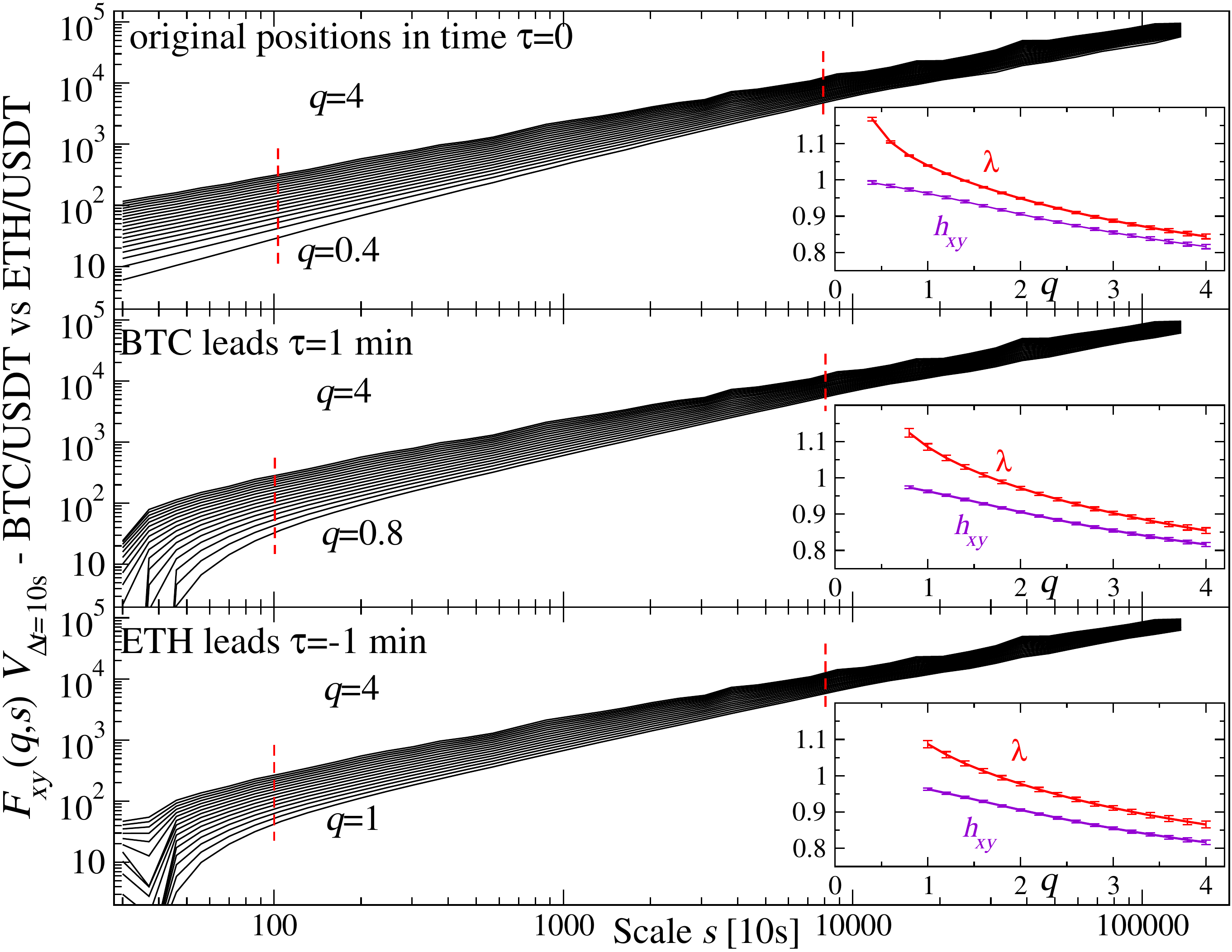}
\caption{(Main panels) Fluctuation functions $F_q^{\rm XY}(s)$ calculated for the time series of volume traded $V_{\Delta t}(t)$ for X=BTC and Y=ETH. In each panel, vertical dashed lines denote a range of time scales $s$ for which a power-law model can be fitted to $F_{\rm XY}^q(s)$. Extreme values of $q$ are also shown. Three cases are considered: both time series are simultaneous (top), time series representing BTC is advanced by $\tau=1$min (middle), and time series representing ETH is advanced by $\tau=1$min. (Insets) The bivariate scaling exponent $\lambda(q)$ vs. the mean univariate scaling exponent $h_{\rm XY}(q)$ calculated for the same time series. Error bars denote the standard errors.}
\label{fig::fluctuation.function.cross.volume}
\end{figure}   

We apply the same formalism to the volume traded $V_{\Delta t}(t)$ and plot the corresponding quantities in Fig.~\ref{fig::fluctuation.function.cross.volume}. There is a slightly broader range of scales for which a power-law dependence can be seen (almost 3 decades for the simultaneous time series and 2 decades for the lagged ones) than it was seen for $r_{\Delta t}(t)$ in Fig.~\ref{fig::fluctuation.function.cross.returns}. The range of $q$ over which multiscaling can be detected is, however, narrower than for $r_{\Delta t}(t)$ in each case, restricted to positive values of $q$ only. Another significant difference concerns $\lambda(q)$ and $h_{\rm XY}(q)$, which are sizeably separated even for $\tau=0$. For $\tau=\pm1$ min this difference is also more pronounced than for the price returns, but here one can notice a broken symmetry between the BTC- and ETH-led time series even in the case of $q=2$: for the advanced ETH the difference $d_{\rm XY}(q)$ is larger than in the opposite case. This effect is also seen in Fig.~\ref{fig::rho.q}(b) (middle). Despite the fact that these gaps are larger, they asymptotically decrease with increasing $q$, suggesting that also in the case of volume $\rho_q(s)$ increases more slowly with $s$ for larger $q$. Examples of this increase of $\rho_q(s)$ are plotted in Fig.~\ref{fig::rho.q}(a) for $q=2$ (middle left) and $q=4$ (middle right). For short scales, the detrended coefficient for the simultaneous time series ($\tau=0$) is substantially larger than for the lagged ones, while for $s>1,000$ (10,000 s) this difference vanishes in both cases (for both $q=2$ and $q=4$). This effect can be understood to be such that, on sufficiently long scales, all the information that has any meaning to the market has already managed to be exchanged by the major cryptocurrencies. On the other hand, the larger values of $\rho_q(s)$ for $\tau=1$ min suggest that information is transferred slightly faster from BTC to ETH than in the opposite direction. This can be explained by the BTC dominance on the market. There is consistency between the respective plots for $\rho_q(s)$ and $d_{\rm XY}(q)$ as regards the growth ratio of the former and the magnitude of the latter for each $\tau$ and each presented $q$.

\begin{figure}[H]
\includegraphics[width=1.0\textwidth]{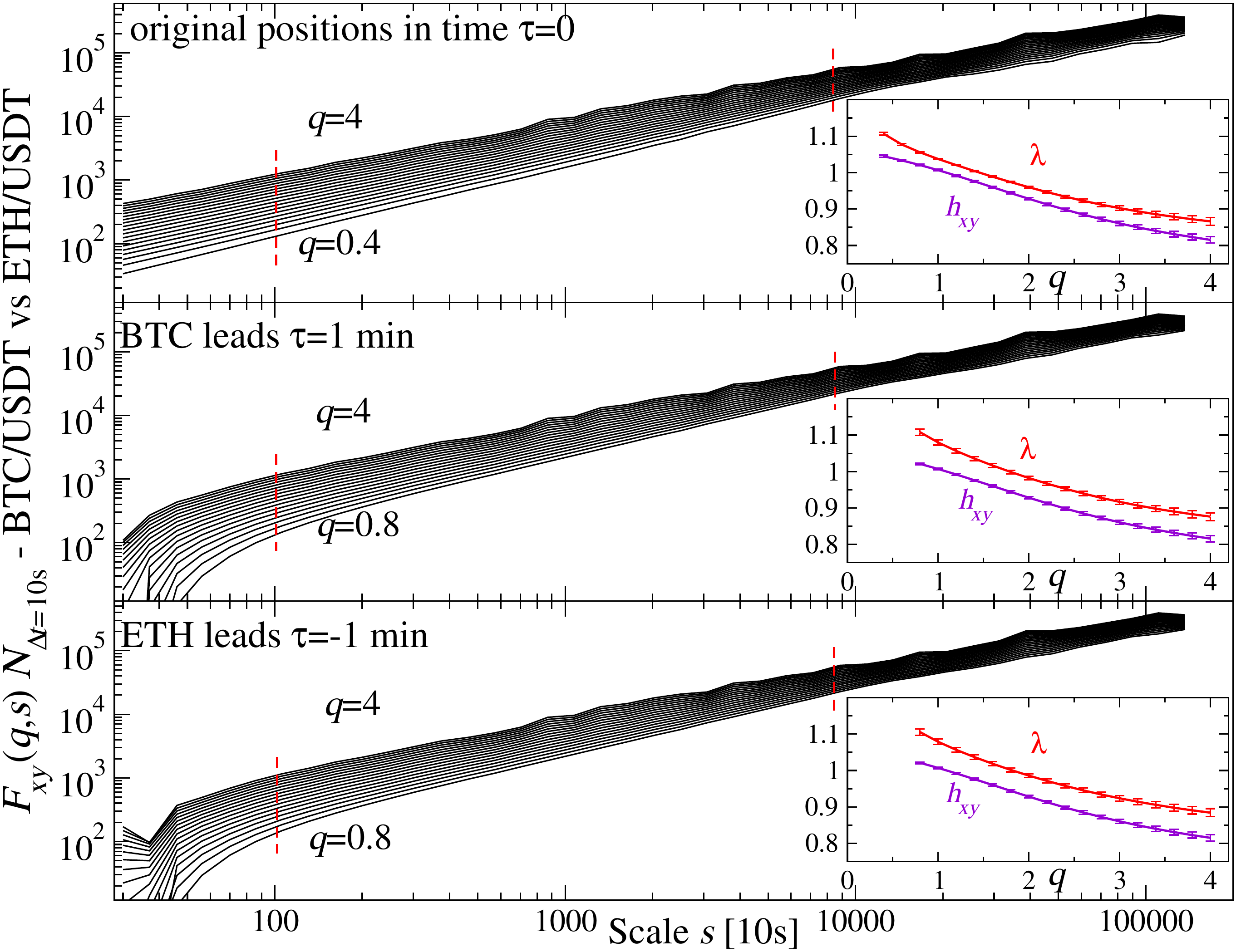}
\caption{(Main panels) Fluctuation functions $F_q^{\rm XY}(s)$ calculated for time series of the number of transactions $N_{\Delta t}(t)$ in $\Delta t$ for X=BTC and Y=ETH. In each panel, vertical dashed lines denote a range of time scales $s$ for which a power-law model can be fitted to $F_q^{\rm XY}(s)$. Extreme values of $q$ are also shown. Three cases are considered: both time series are simultaneous (top), time series representing BTC is advanced by $\tau=1$min (middle), and time series representing ETH is advanced by $\tau=1$min. (Insets) The bivariate scaling exponent $\lambda(q)$ vs. the mean univariate scaling exponent $h_{\rm XY}(q)$ calculated for the same time series. Error bars denote the standard errors.}
\label{fig::fluctuation.function.cross.transactions}
\end{figure}   

The behaviour of $F_q^{\rm XY}(s)$ for the time series of the mean number of transactions $N_{\Delta t}(t)$ resembles that for $V_{\Delta t}(t)$ - see Fig.~\ref{fig::fluctuation.function.cross.transactions} (main plots). What is different is the behaviour of the exponents $\lambda(q)$ and $h_{\rm XY}(q)$, because while we increase $q$, we observe that they are getting closer to each other first, then somewhere for $1.5 < q < 2$ they reach a minimum distance and start to diverge for larger $q$s (insets in Fig.~\ref{fig::fluctuation.function.cross.transactions}). This is even better visible in their difference in Fig.~\ref{fig::rho.q}(b) (bottom), which is associated with a moderate increase of $\rho_q(s)$ for $q=2$ (bottom left) of Fig.~\ref{fig::fluctuation.function.cross.transactions}(a) and more a pronounced increase for $q=4$ (bottom right). Like in the case of volume, up to a scale of $s\approx 1000$ (i.e., 10000 s) the coefficient is higher for the simultaneous signals than it is for the shifted ones.

\section{Discussion and conclusions}
\label{sect::discussion}

In this work, we studied the detrended cross-correlations between three trading characteristics of two major cryptocurrencies: BTC and ETH over the last 2 years. This period was characterised by the profound stress imposed on the world economy by the Covid-19 pandemic, but already after the initial shock of early 2020, when most of the markets experienced heavy losses, there was a quick rebound to new all-time highs in the second half of the year. During the pandemic, both the traditional markets and the cryptocurrency market passed through different stages alternately governed by strong ``bears'' and ``bulls''. Recently, the military escalation in Ukraine has also been exerting a heavy impact on the markets, including the cryptocurrency market, leading them to suffer from extra draw-downs that added momentum to the bear market dominating the scene since December 2021. From the cryptocurrency perspective, another interesting structural phenomenon is an emergent, strong permanent coupling of the cryptocurrency market and the stock market~\cite{aslanidis2021,watorek2021,kwapien2021,james2021c,james2021d} -- a phenomenon that prior to the pandemics used to be observed only occasionally~\cite{drozdz2020a,drozdz2020b,conlon2020,corbet2020,demir2020,kristoufek2020,james2022}.

Financial market data is well-known of its nonstationarity. This is why the classical approach to quantifying correlations among such data in terms of, for instance, the Pearson coefficient can overlook important properties of the data. The formalism based on the detrended fluctuation analysis~\cite{peng1994} and its further generalisations~\cite{kantelhardt2002,podobnik2008,zhou2008,oswiecimka2014} are much more suited for non-stationary signals, as their inherent feature is the elimination of multiscale trends. At present, this approach is favoured and becomes a standard tool of time series analysis. In our work, this formalism was employed to investigate cross-correlations between two major cryptocurrencies -- BTC and ETH. We selected three trading characteristics: price returns, volume, and number of transactions in time unit (chosen to be 10s). We did not restrict the analysis to the time series collated parallel in time, but we also analysed the ones that were shifted relative to each other by 1 min. By looking at the univariate and bivariate fluctuation functions, we found that both were manifesting the multifractal property. Only the results for the price returns can be directly compared with the analogous results obtained for the data covering earlier periods before Covid-19~\cite{takaishi2018,takaishi2019,Han2020,kristjanpoller2020,bariviera2021,watorek2021}. In this case, we see that the fractal properties of the price returns have not changed much. This is true not only if we look at the fluctuation functions, but also at the bivariate and mean univariate scaling exponents, which are almost equal in this case.

We also analysed for the first time in the literature the lagged cross-correlations between the time series representing BTC and ETH. We found that shifting one of the time series suppresses the cross-correlation magnitude, but nevertheless the remaining cross-correlations are still significant, especially for the scales larger than a few hours, during which the information that arrives on the market is able to disperse. The time series of the price returns did not reveal any asymmetry between the cases, in which either BTC or ETH is lagged, for $q=2$. However, the time series of the volume traded and the number of transactions showed a small effect of such an asymmetry: if BTC led, the cross-correlation was slightly stronger, which we interpreted as an effect of faster information transfer in the direction BTC$\rightarrow$ETH than in the opposite direction that we related to the BTC dominance on the cryptocurrency market. Interestingly, for $q=4$ this asymmetry is easy to notice in the case of the returns as well -- manifestation of a fact that the correlations associated with the returns of large amplitude are more direction-sensitive than the correlations associated with the small and medium returns. Another observation was that, for short scales, the strength of the cross-correlations between the simultaneous time series representing BTC and ETH was the largest for the price returns, whereas for the volume and the number of transactions it was substantially smaller. This effect was less pronounced for the lagged time series and it gradually disappeared for the longer scales.

A general conclusion that we can draw from this study is that the overall fractal properties of the major-cryptocurrency time series are stable as regards the pre-Covid, Covid, and post-Covid periods. We can interpret this stability as related to the fact that the processes that govern the fractal organisation of the cryptocurrency trading data are so fundamental that they can resist the social and economical forces perturbing the market. This conclusion apparently challenges some earlier reports based on the stock market data, whose fractal properties were market-phase dependent (e.g.,~\cite{grech2010}), but actually it has to be noticed that here we did not attempt to decompose the market dynamics and presented the average market properties only. As the fractality indicates that the processes driving the cross-asset trading characteristics are largely scale-invariant, our result can be of significance from the market-modelling perspective. It also provides further support to a thesis that the cryptocurrency market has reached maturity at least from the statistical and dynamical points of view.

It should be noted, however, that our study has its inherent limitations. We analysed two major cryptocurrencies only, it is thus conceivable that results for other cryptoassets, especially those with much a worse liquidity, would potentially diverge from the results presented here. We also stress that we restricted our analysis to a few values of time lag. If another study considered different lags, especially the shorter ones, both the the cross-correlation magnitude and the asymmetry effect could be different. We also analysed a particular time period only, while the history of the cryptocurrency market is much longer. We could thus expect that the properties of the lagged cross-correlations evolve in time as well. Each of these limitations should be overcome in future research based on extended data. Another issue to be considered in future is to focus on short periods of time in order to gain some insight into how the fractal cross-correlation magnitude and lag asymmetry depend on the market situation (the bull and bear markets, for instance).

\vspace{6pt} 

\authorcontributions{Conceptualization, S.D., J.K. and M.W.; methodology, S.D., J.K. and M.W.; software, M.W.; validation, S.D., J.K. and M.W.; formal analysis, M.W.; investigation, S.D., J.K. and M.W.; resources, M.W.; data curation, M.W.; writing---original draft preparation, J.K.; writing---review and editing, S.D., J.K. and M.W.; visualization, M.W.; supervision, S.D. All authors have read and agreed to the published version of the manuscript.}

\funding{This research received no external funding.}

\dataavailability{Data available freely from Binance~\cite{binance}.} 

\begin{adjustwidth}{-\extralength}{0cm}

\reftitle{References}

\end{adjustwidth}

\end{document}